\documentclass[pre,nofootinbib,twocolumn,showpacs,preprintnumbers,amsmath,amssymb,floatfix,superscriptaddress]{revtex4}
\usepackage{amsmath}
\usepackage{amsfonts}
\usepackage{dcolumn}                    	     
\usepackage{amssymb}
\usepackage{graphicx,graphics,wrapfig,rotating}     	
\usepackage[german,english]{babel}	
\usepackage{bm,fancybox}                	     

\def\bra#1{{\langle#1|}}
\def\ket#1{{|#1\rangle}}

\def\expect#1{{\langle#1\rangle}}
\def\e{{\rm e}}

\def\Zhat{{\hat Z}}

\def\x{{\hat x}}

\begin{document}

\title{Behavior of the current in the asymmetric quantum multibaker map}

\author{Leonardo Ermann}
\affiliation{Departamento de F\'\i sica, CNEA, Libertador 8250, (C1429BNP) Buenos Aires, Argentina}
\affiliation{Departamento de F\'\i sica, FCEyN, UBA, Pabell\'on 1 Ciudad 
Universitaria, C1428EGA Buenos Aires, Argentina}
\author{Gabriel G. Carlo} 
\affiliation{Departamento de F\'\i sica, CNEA, Libertador 8250, (C1429BNP) Buenos Aires, Argentina}
\author{Marcos Saraceno} 
\affiliation{Departamento de F\'\i sica, CNEA, Libertador 8250, (C1429BNP) Buenos Aires, Argentina}
\affiliation{Escuela de Ciencia y Tecnolog\'\i a, UNSAM, Alem 3901, B1653HIM Villa Ballester, Argentina}

\date{\today}

\pacs{05.45.Mt, 05.40.Jc, 05.60.-k}

\begin{abstract}

Recently, a new mechanism leading to purely quantum directed transport in the asymmetric 
multibaker map has been presented. Here, we show a comprehensive characterization of 
the finite asymptotic current behavior with respect to the $h$ value, the shape of the initial 
conditions, and the 
features of the spectrum. We have considered different degrees of asymmetry in these studies and we 
have also analyzed the classical and quantum phase space distributions for short times in order 
to understand the mechanisms behind the generation of the directed current. 
\end{abstract}

\maketitle

\section{Introduction}

There is a great interest in the study of directed transport in unbiased periodic systems. 
This phenomenon, also referred to as the ratchet effect, was initially considered by Feynman \cite{Feynman}. 
It can be classically ascribed to breaking all spatiotemporal symmetries leading to 
momentum inversion \cite{origin}. This allows a net current generation. For example, in non-Hamiltonian 
systems chaotic attractors need to be asymmetric \cite{nonHam} 
whereas in Hamiltonian ones (with mixed phase spaces) a chaotic layer should have this property \cite{Ham}. 
Many times the same principle translates almost directly into the quantum domain \cite{asymFloquet}, but 
in other cases more complex behaviors arise \cite{Qeffects}. 

Since the first studies the relevance of this subject has been steadily growing, and several 
fundamental questions about the origin and properties of the net current have been answered \cite{Reimann}. 
However, the considerable amount of possible applications have opened a very broad field of research. In fact, 
a great and increasing number of experiments implement different kinds of ratchets. In biology, 
molecular motors principles can be understood on these grounds \cite{biology}. 
Also, they can be useful to develop nanodevices like rectifiers, pumps, particle separators, molecular 
switches and transistors \cite{nanodevices}. Cold atoms and Bose-Einstein condensates have emerged 
as a very active area of application of these ideas, and the first experiments 
have initiated an activity that continues until present \cite{CAexp}. These efforts have led to the 
very recent success in transporting Bose-Einstein condensates for particular initial conditions 
by relying on purely quantum ratchet accelerators mechanisms \cite{BECratchets}. Such experiments 
involve essentially the atom optics kicked rotor
\cite{AOKR} at quantum resonance. In this system the current has no 
classical analogue and can be generated by just breaking the spatial symmetry \cite{purelyQR}.
Though the experimental realization of some proposed models is still demanding and the 
theoretical explanations are still not complete, ongoing studies show several new proposals 
\cite{recentStudies}. They include ways of coherently controlling the ballistic energy growth 
of the atoms \cite{coherentControl}.

In order to investigate the mechanisms leading to net transport generation in quantum systems we 
have recently introduced an asymmetric version of the quantum multibaker map that shows a finite asymptotic 
current with no 
classical counterpart \cite{previousPaper}. This is a paradigmatic model in classical and quantum chaos, 
but also in statistical mechanics \cite{modelSource1,modelSource2}. 
In this work we study the properties of the directed current in depth. 
We provide with a characterization of its behavior as a function of 
the $h$ value, the initial conditions and the spectrum features. All this has been considered for 
different values of the main parameter which determines the degree of spatial asymmetry. 
With this results at hand we proceed further to study the classical and quantum versions of the phase space distributions 
for short times. This shows the way in which the quantum current arises and the classical one does not, providing 
with a firm ground in order to understand the mechanisms involved. We finally make 
a comparison with the behavior of the system for longer evolutions of the order 
of the Heisenberg time.

In the following we describe the organization of this paper. In Section II we present our model in detail and the methods 
we have used to study it. We have chosen to divide this Section in four parts. Firstly, we formulate the classical 
and quantum propagators, then we explain some properties of the second one that are useful for the time evolution. 
Also, we introduce an asymptotic expression for the coarse-grained current, which is the main quantity under 
investigation. Finally the symmetry properties are explained. In Section III we 
analyze the current behavior as a function of 
$h$, the initial conditions, and the spectrum shape. In Section IV we show 
the connection between the symmetries and the current generation by focusing on the classical 
and quantum phase space distributions for short times. We establish 
how the degree of asymmetry influences the features of the system studied in the previous 
Section. Finally, Section V is devoted to the conclusions. 

\section{Model and Methods}
\subsection{Classical and quantum propagators}

\begin{figure}[htp]
 \includegraphics[width=0.47\textwidth]{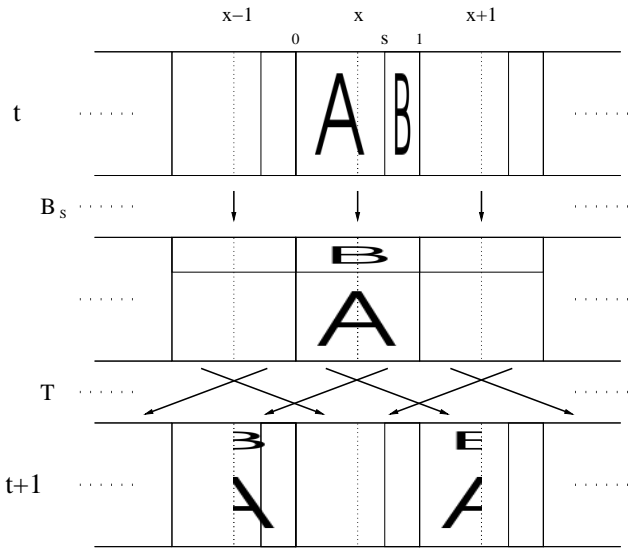}
\caption{Geometric action of the asymmetric multibaker map. One iteration of the map corresponds to 
a composition of an internal evolution (given by the asymmetric baker map), and a translation 
among adjacent cells (which depends on the position inside of them).}
\label{fig:bakerasim}
\end{figure}

The classical multibaker map \cite{modelSource1} is defined in a phase space consisting of a lattice of 
unit square cells in position direction and confined in momentum ($p\in[0,1)$). A phase space point can be 
completely defined by the number $x$ ($x\in\mathbb{Z}$) of the cell to which it belongs and the position 
and momentum inside of it ($q,p\in[0,1)$). The action of the map is a composition of an internal evolution 
inside of each cell (the baker map), followed by a translation along the lattice given by
\begin{equation}
\label{eq:Cmultibak}
 M_{s}=T \circ B_{s}. 
\end{equation}
In this expression $B_s$ is the asymmetric baker's map in the unit square phase space cell $x$. This is 
the area preserving map
\begin{equation}
\label{eq:bakerasimGeneral}
 B_{s}(q,p)\equiv\left\lbrace \begin{array}{cc} \left( \frac{1}{s}q, s p\right) 
&0\leq q < s\\ 
\left((1-s)^{-1} (q-s),(1-s) p +s \right) &s\leq q< 1
\end{array}
 \right. 
\end{equation}
It can be clearly seen that the degree of asymmetry is controlled by the parameter $s$, and that 
there are two different Lyapunov exponents $\lambda_{1}=-\ln{(s)}$, $\lambda_{1}=-\ln{(1-s)}$.
On the other hand $T$ corresponds to an unbiased translation along the lattice, defined by 
\begin{equation}
 T=\left\lbrace \begin{array}{cc}\left( x+1,q,p\right) &0\leq q < 1/2\\
 \left( x-1,q,p\right) &1/2\leq q < 1
\end{array}
\right. 
\end{equation}
This translation can only occur among adjacent cells and depends on the position inside of them.
The geometric action of the asymmetric multibaker map (AMBM) can be seen in Fig. \ref{fig:bakerasim}.
The asymmetric quantum multibaker map (AQMBM) is defined in a Hilbert space $\mathcal{H}$ which is 
the direct product of the lattice space ($\mathcal{H}_L$), and the individual cell space 
($\mathcal{H}_B$), $\mathcal{H}=\mathcal{H}_L\otimes\mathcal{H}_B$ \cite{modelSource2,leo1}. 
In this work we will consider even $D$-dimensional internal subspaces $\mathcal{H}_B$ on a torus (where 
$h=1/D$), and 
infinite dimensional lattice subspaces.
The translation over the lattice will be similar to the classical one. The dependence on 
the position inside of each cell is now given by the unbiased projectors $\hat{P}_R$ and $\hat{P}_L$. 
These operators perform the projection on the right and left half of the position basis inside 
of each cell, satisfying $\hat{P}_R + \hat{P}_L = \hat{I}$ and ${\rm Tr}\left(\hat{P}_R\right)={\rm Tr}
\left( \hat{P}_L\right) = D/2$. Therefore, the AQMBM can be written as 
\begin{equation}
\hat{M}_{s} \equiv  \hat{T}  \circ \hat{B}_{s}= 
\left( \hat{U} \otimes\hat{P}_{R} + 
\hat{U}^{\dagger} \otimes \hat{P}_{L}\right) \left( \hat{I} \otimes \hat{B}_{s} \right)
\label{eq:Qmultibak}
\end{equation}
where  $\hat{U}$ is a unitary translation operator acting on the lattice subspace 
$\hat{U} |x\rangle = |x+1\rangle$ (with $\{|x\rangle, x=\ldots,-2,-1,0,1,2,\ldots\}$ 
taken as the position basis set of the lattice). $\hat{B}_{s}$ is   
\begin{eqnarray}
 \hat{B}_{s}&=&\hat{G}^{\dagger}_{D}\left(\begin{array}{cc}\hat{G}_{D_{1}}&0\\0&
\hat{G}_{D_{2}}\end{array}\right)\\
\left(\hat{G}_{D}\right)_{kl}&\equiv& D^{-1/2}e^{-i 2\pi(k+
1/2)(l+1/2)/D}. 
\end{eqnarray}
This is the asymmetric quantum baker's map with antiperiodic boundary conditions, i.e., 
the corresponding generalization of the quantum symmetric one \cite{voros,saraceno}. 
In this case only the values of $s$ such that $D_{1}=sD$ and $D_{2}=D-D_{1}$ are 
 positive integer numbers are allowed. 

\subsection{Time evolution}

The time evolution of an initial state can be computed straightforwardly in both classical 
and quantum cases in terms of the propagators given in Eq. (\ref{eq:Cmultibak}) and Eq. (\ref{eq:Qmultibak}), 
respectively. As usually happens in directed transport studies we are interested in the behavior 
of an initially localized distribution of particles. For that reason, we will focus on initial states 
which are located in a single site of the lattice. In the classical case the initial state will be a 
a uniform probability distribution with the shape of a momentum band of width $\delta p$ and extending 
completely along the $q$ coordinate of the initial cell. 

Correspondingly, in the quantum case we will always start with separable initial states of the form  $\rho_{0}=\rho_0^L\otimes\rho_0^B$. In this case, $\rho_0^L$ is the initial state in the lattice 
space, in practice a given position basis element. On the other hand, $\rho_0^B$ is a mixed superposition 
of $\Delta p$ momentum eigenstates of the individual cell subspace. This kind of initial state 
is the quantum analogue of the previously described classical one, therefore we will take 
$\Delta p=D\delta p$ to make both of them fully comparable.

The quantum state at time $t$, $\rho(t)$, is the result of the discrete time propagation of 
the initial state given by 
\begin{equation}
\rho(t)=\left(\hat{M}_{s} \right)^t\rho_0\left(\hat{M}_{s}^{\dagger} \right)^{t}.
\label{qevolution}
\end{equation}
This expression can be simplified noting that in $\hat{M}_{s}$ the translation operator $\hat{U}$ 
becomes diagonal in the momentum basis of the lattice subspace $\{|k\rangle\}$
\begin{equation}
\hat{U}\vert k\rangle = e^{-ik}\vert k\rangle
\end{equation}
where by the previous definition
\begin{equation}
\vert k\rangle = \sum_{x=-\infty}^{\infty} \vert x\rangle e^{ikx}.
\end{equation}
Thanks to this property we can better handle the action of the 
AQMBM of Eq. (\ref{eq:Qmultibak}) on a given 
state of our system. If we define $\hat{B}_{s,k}$ as an operator acting on individual cell 
states $\vert\Psi_B\rangle$ and being parametrized by the lattice momentum value $k$, $\hat{M}_{s}$ 
can be rewritten as
\begin{equation}
 \hat{M}_{s} \left(\vert k\rangle\otimes\vert\Psi_B\rangle\right) =  \vert k\rangle\otimes\hat{B}_{s,k} 
\vert\Psi_B\rangle,
\end{equation}
where by definition
\begin{equation}
\hat{B}_{s,k} \equiv \left( \begin{array}{cc} e^{-ik} & 0 \\ 0 & e^{ik}   \end{array}\right) \hat{B}_{s}
\end{equation}
Then, the quantum asymptotic time evolution turns into the study of the eigenvalues 
and eigenvectors of this last operator, as we will see in the following.

\subsection{Coarse-grained current}

For a given ensemble of classical initial conditions, we define $p_{class}(x,t)$ as the 
probability of the particle to be in the $x$ lattice cell at time $t$. In this way we can  
compute the mean value of the coarse-grained position as $\langle x \rangle = \sum_x x 
\, p_{class}(x,t)$ (which is the average value of the cell position $x$). 
Then, the coarse-grained current is calculated as the difference between 
this mean value at time $t$ and the same value taken at an earlier time $t-1$. 
The current $J_{class}=\langle x(t)\rangle-\langle x(t-1) \rangle$ can 
be derived from the first moment of the classical distribution, but higher moments can be calculated 
also in this way, i.e. disregarding the fluctuations that take place inside each cell. 

For the quantum evaluation we first consider the probability distribution of the particle 
to be in the $x$ lattice cell after $t$ iterations of the map. This is given by
\begin{equation}
p(x,t)={\rm Tr}\left[ \rho(t) \left(\vert x\rangle\langle x\vert \otimes I\right) \right]
\end{equation}
In particular, for an initial state localized in one site 
(i.e., for which we take $\rho_0^\text{L}=\vert0\rangle\langle 0 \vert$) and in the lattice momentum 
representation, the previous expression becomes
\begin{equation}
p(x,t)= \int \int \frac{dk dk'}{(2\pi)^2}   \e^{-ix(k-k')}
   {\rm Tr}\left[\left(\hat{B}_{s,k'}\right)^t \rho^\text{B}_0 \left(\hat{B}^\dagger_{s,k}\right)^t  \right] 
\end{equation}
The coarse grained position is obtained by tracing out each cell's internal degrees of freedom ($q$). 
The moments of this quantity can now be easily calculated using the probability distribution $p(x,t)$ 
\begin{equation}\label{eq:moments}
 \langle x^m\rangle_t=\sum_x x^m p(x,t).
\end{equation}
Finally, in complete analogy to the classical definition we will take the quantum 
coarse-grained current to be
\begin{equation}
 J(t)=\langle x\rangle_t-\langle x\rangle_{t-1} \label{eq:current}.
\end{equation}

Following closely Brun \textit{et al.} \cite{Brun}, we insert the identity 
\begin{equation}
\frac{1}{2\pi} \sum_x x^m \e^{-ix(k-k')}
  = i^m \delta^{(m)}(k-k')
\end{equation}
into Eq. (\ref{eq:moments}), and integrating by parts we obtain
\begin{equation}
\langle x^m\rangle_t = \frac{i^m}{2\pi} \int dk \
 {\rm Tr}\left[\rho^\text{B}_0 \left(\hat{B}^\dagger_{s,k}\right)^t \frac{d^m}{dk^m}\left(\hat{B}_{s,k}\right)^t  \right]. 
\end{equation}
Therefore the first moment can be written as
\begin{equation}
\expect{\x}_t = \frac{i}{2\pi} \int dk
 \ {\rm Tr}\left[\rho^\text{B}_0\left(\hat{B}^\dagger_{s,k}\right)^t \left( \frac{d}{dk}
  \left(\hat{B}_{s,k}\right)^t \right) \right],
\label{firstmoment1}
\end{equation}
where
\begin{eqnarray}
\frac{d\hat{B}_{s,k}}{dk}&=&\left(-i\e^{-ik}\hat{P}_R+i\e^{ik}\hat{P}_L\right)\hat{B}_{s}=-i\hat{Z}\hat{B}_{s,k},
\,\,{\rm and}\nonumber\\
\hat{Z}&\equiv&\hat{P}_R-\hat{P}_L. 
\end{eqnarray}

Substituting this into Eq. (\ref{firstmoment1}), the coarse-grained position mean value becomes 
\begin{equation}
\expect{\x}_t=\sum_{j=1}^t \int \frac{dk}{2\pi}
 \ {\rm Tr}\left[\rho_0^\text{B} \left(\hat{B}_{s,k}^\dagger\right)^j \Zhat
  \left(\hat{B}_{s,k}\right)^j\right].
\label{firstmoment2}
\end{equation}
A similar procedure could be followed to obtain higher moments.

The time dependence in Eq. (\ref{firstmoment2}) can be made explicit by considering 
the spectral properties of the map $\hat{B}_{s,k}$ 
\begin{equation} 
\hat{B}_{s,k} \ket{\phi_{l}(k)}=\exp(i\theta_{l}(k)) \ket{\phi_{l}(k)}.
\label{spectqmap}
\end{equation}
In this basis the initial cell distribution is 
\begin{equation}
 \rho_0^\text{B}=\sum_{ll^\prime} a_{ll'}(k) \vert\phi_{l}(k)\rangle\langle\phi_{l^\prime}(k)\vert.
\end{equation}
Substituting this into Eq. (\ref{firstmoment2}) for the first moment we obtain
\begin{equation}
\label{firstmoment3}
\expect{x}_t = \int \frac{dk}{\pi} \sum_{l,l'}
 a_{ll'}(k) \bra{\phi_{l}(k)} \hat{Z} \ket{\phi_{l'}(k)} 
  \sum_{j=1}^t \e^{i(\theta_{l'}(k)-\theta_{l}(k))j} \;.
\end{equation}

No approximations have been made in this derivation. If the spectrum has no degeneracies, as 
will be the case for chaotic maps most of the terms in Eq. (\ref{firstmoment3}) will be highly oscillatory; 
hence, over time, they will average to zero. Only the diagonal terms in the above sum
are nonoscillatory, allowing us to write
\begin{equation}
\langle x\rangle_t = J_\infty t + \text{oscillatory terms},
\end{equation}
where
\begin{eqnarray}
\label{firstmoment4}
J_\infty &=& \int \frac{dk}{2\pi} \sum_l a_{ll}(k) Z_{ll}(k)\\ 
Z_{ll^\prime}(k)&\equiv&\langle \phi_{l}(k)\vert \hat{Z} \vert\phi_{l^\prime}(k)\rangle. 
\end{eqnarray}
In these expressions, $J_\infty$ is the asymptotic value of the coarse-grained current defined in 
Eq. (\ref{eq:current}). The quantity $a_{ll}(k)$ corresponds to the projection of the initial 
state in the basis of eigenstates as previously stated, 
and $Z_{ll}(k)$ is a kind of right-left balance of each eigenstate.

This completes the description of the methods used to study our system. In the following we will 
explain some symmetry considerations relevant for the directed transport mechanism.

\subsection{Symmetry properties}

By looking at Fig. \ref{fig:bakerasim} the first thing that can be seen is that, 
though the baker map we consider is asymmetric, the transport term is 
unbiased. The transport is only due to this translation, that maps the same volume 
of phase space to the right and left. Quantum mechanically this also means 
that there is no tunneling effects from cell to cell. 
It has been shown that the presence of the net classical transport is originated 
from breaking all spatiotemporal symmetries that leave the system unchanged 
but change the sign of the (coarse-grained) current \cite{origin}. 
There are two transformations that fulfill these conditions, 
let us consider first
\[
S_I: \, \, \, q \rightarrow 1-q; \, \, \, p \rightarrow 1-p,
\] 
acting on each cell, and leaving the transport term $T$ unchanged. 
Under the action of $S_I$, the $q$ and $p$ coordinates are reflected 
with respect to their midpoints at each cell, and the map $B_{s}$ transforms 
to $B_{1-s}$ (we underline that this is valid in the classical and 
in the quantum case).
For $s=1/2$, i.e. the symmetrical Baker map, this transformation is a 
symmetry of the system. But it also changes the sign of the coarse-grained 
current, since a given trajectory that is transported to the left (right) 
at each iteration is now transported to the right (left). For other 
values of $s$ the symmetry is broken.
The other transformation is
\[
S_{II}: \, \, \, q \rightarrow p; \, \, \, p \rightarrow q; 
\, \, \, T \rightarrow T^{-1}; \, \, \, t \rightarrow -t, 
\]  
where the $q$ and $p$ part acts on each cell. This is the 
time reversal symmetry, present for any value 
of $s$. This transformation 
leaves the system unchanged, but reverses all trajectories and 
consequently changes the sign of the coarse-grained current. 
This forbids any classical current for unbiased initial conditions. 
In previous studies we have found transient effects for biased conditions but they 
disappear very rapidly due to the exponential mixing property of 
the Baker map.

Finally we will refer to the symmetry properties of the coarse-grained current.
$J$ is an odd function of $s$ around $s=0.5$, i.e. 
$\langle J_{s}\rangle=-\langle J_{1-s}\rangle$. 
In fact, if we apply the symmetry transformation $S_I$ to Eq. (\ref{qevolution}), 
and then trace out the internal degrees of freedom inside of each cell we 
obtain that $p_s(x,t)=p_{1-s}(-x,t)$ for all $t$. This result is valid for 
any initial $ \rho_0^\text{B}$ symmetrical under $S_I$.

\section{Quantum current behavior}

\begin{figure}[htp]
\begin{center}
 \includegraphics[width=0.47\textwidth]{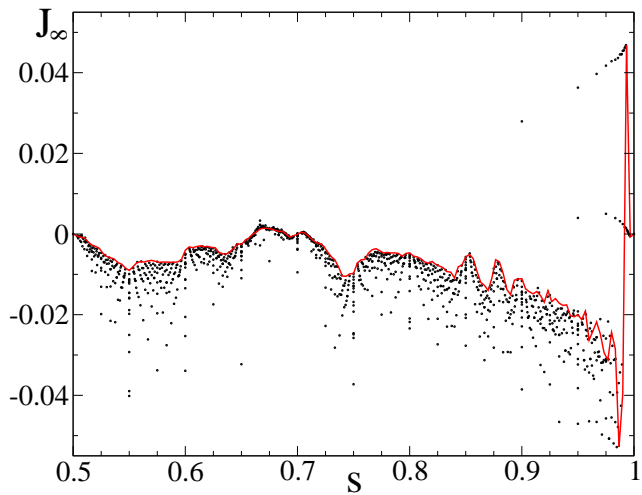}
\caption{(Color online) Asymptotic coarse-grained current $ J_\infty$ 
for the AQMBM as a function of $s=D_1/D$, for all possibles values of $D_1\ge D/2$. 
The asymptotic current is represented by a solid line for $D=300$ and with 
dots for lower values. The initial state is an equal probability mixture 
of $\Delta p=D/10$ central momentum eigenstates of the cell.}
\label{fig:c1}
\end{center}
\end{figure}

In this Section we analyze the most important aspects of the quantum directed current, 
providing with a comprehensive understanding of its behavior. 
In the first place, we study the transition towards the classical limit that 
allows us to see the way in which the net transport vanishes. For that purpose 
we have numerically evaluated the asymptotic value of the coarse-grained 
quantum current $ J_\infty$ by means of Eq. (\ref{firstmoment4}). 
This has been done for all possible values of the quantum asymmetry parameter $s=D_1/D$, 
taking only $s \ge 0.5$ thanks to the symmetry property explained in the Section II. 
In order to have the same classical limit for all the $h=1/D$ values, we have taken as 
initial conditions equal probability mixtures of an (integer) number $\Delta p=D/10$ 
of central momentum eigenstates. 
The results can be seen in Fig. \ref{fig:c1}, where the solid line corresponds to a dimension $D=300$ 
for the Hilbert space of the cell, and the dots coorespond to all possibles values of $D$ 
which are divisible by $10$, between $D=20$ and $D=290$.

We can see that the currents corresponding to $D_1=D-1$ and $D_1=D-2$ 
are clearly different from the general behavior, we will come  
back to this particular feature later on when we analyze the spectrum. However, we 
note that there is a global convergence to the solid line, though the dependence 
on $s$ is rather non-trivial. In fact, the current behavior (with the exception 
of the last points for $D_1=D-1$ and $D_1=D-2$) can be divided into two parts. 
The first one corresponds to $s\lesssim0.7$, where $J_\infty$ is already small for the 
maximum $D$ we have taken in our calculations. In this respect, the current seems to 
vanish much faster than in the $s\gtrsim0.7$ domain, in which higher values can be 
observed. It seems that the quantum effects are enhanced if one of the two parts 
in which the phase space is divided is clearly smaller than the other. We have found 
a similar effect in our studies of the current dependence on the initial conditions. 
For that reason we pay special attention to these cases in the last part of 
this Section.

\begin{figure}[htp]
\begin{center}
 \includegraphics[width=0.47\textwidth]{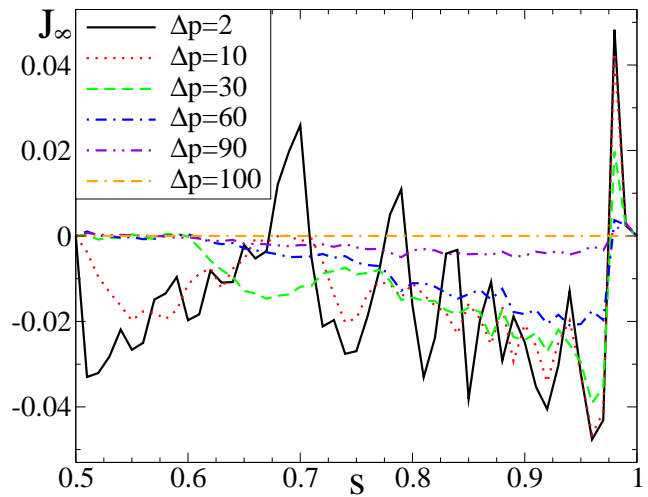}
\caption{(Color online) Asymptotic coarse-grained current $J_\infty$ 
for the AQMBM as a function of $s=D_1/D$ and for a fixed $D=100$. 
The initial states and values of $s$ are taken as 
in Fig. \ref{fig:c1}, but for $\Delta p$ equal to $2$, $10$, $30$, $60$, $90$ 
and $100$.}
\label{fig:initial}
\end{center}
\end{figure}

We have also focused on the behavior of the asymptotic coarse-grained current as a 
function of the width in $p$ of the initial mixed superposition of momentum eigenstates 
The values of $ J_\infty$ for a fixed dimension $D=100$, different $\Delta p$ and 
as a function of $s$, can be seen in Fig. \ref{fig:initial}. 
The current decreases with the width of the 
momentum band in the region of $s\lesssim0.7$. Nevertheless, for $s\gtrsim0.7$ 
we can see that by enlarging the width of the initial distribution up to approximately 
a 60\% of the maximum phase space size in momentum, the fluctuations become smoother. 
However, it is remarkable that the current nearly vanishes in the same region where 
the convergence to the classical behavior is faster. For greater $\Delta p$ values 
the current decreases strongly, and for a distribution over all the $p$ range 
there is no current. 

\begin{figure}[htp]
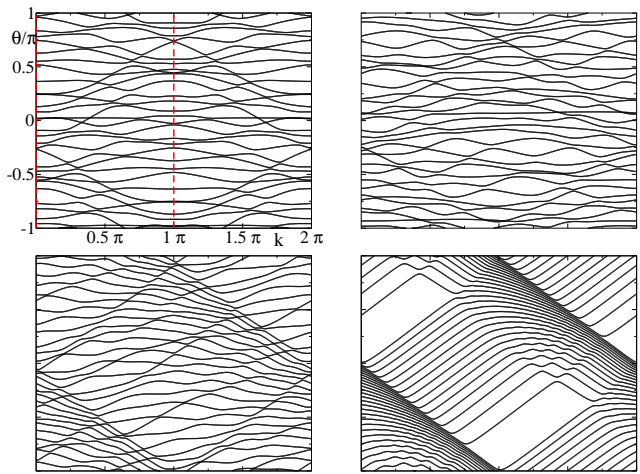

\begin{center}
 \includegraphics[width=0.235\textwidth]{4a.eps}
 \includegraphics[width=0.235\textwidth]{4b.eps}\\
 \includegraphics[width=0.235\textwidth]{4c.eps}
 \includegraphics[width=0.235\textwidth]{4d.eps}
\caption{Eigenphases $\theta$ (in units of $\pi$) of the AQMBM as a 
function of $k$ for $D=30$. On the top panels 
we can find them for $D_1=15$ (left) and $D_1=16$ (right), and on the bottom ones for 
$D_1=26$ (left) and $D_1=29$ (right). For $D_1=15=D/2$, the spectrum has 
a reflection symmetry at $k=\pi$ and $k=0$ (both indicated with 
red dashed lines).
}
\label{fig:specD30dk}
\end{center}
\end{figure}

Finally, in view of the relevance that the operator $\hat{B}_{s,k}$ has in the properties 
of $J_\infty$, we have studied some features of its spectrum for different values of $s$. 
We display the eigenphases $\theta$ (in units of $\pi$) as a function of $k$ 
in Fig. \ref{fig:specD30dk}, for 
$D=30$. The spectrum for the case $D_1=15$, for which the symmetry $S_I$ is 
present, is invariant under reflections at $k=\pi$. This is due to the 
fact that $\hat{B}_{s,k}$ is invariant under $k \rightarrow 2\pi-k$, 
up to an even number of row permutations. The periodicity 
in $k$ makes the spectrum symmetric with respect to $k=0$ also.
This symmetry is absent for all the other values of $s$. We have considered 
the less asymmetric case $D_1=16$, and an intermediate one with $D_1=26$, 
where this becomes already evident. Finally, for $D_1=29$ we can see a very 
regular spectrum, similar to those of integrable systems, 
that nevertheless shows level repulsion. In all cases, there is 
a symmetry given by the transformation $k \rightarrow k+\pi$, 
$\theta \rightarrow \theta+\pi$ since $\hat{B}_{s,k+\pi}=-\hat{B}_{s,k}$, 
and therefore any eigenstate of $\hat{B}_{s,k+\pi}$ ($\vert\phi_{l}(k+\pi) \rangle$) will 
be also an eigenstates of $\hat{B}_{s,k}$ with eigenvalue $\theta_{kl}=\theta_{k+\pi l}+\pi$.

\begin{figure}[htp]
\begin{center}
 \includegraphics[width=0.47\textwidth]{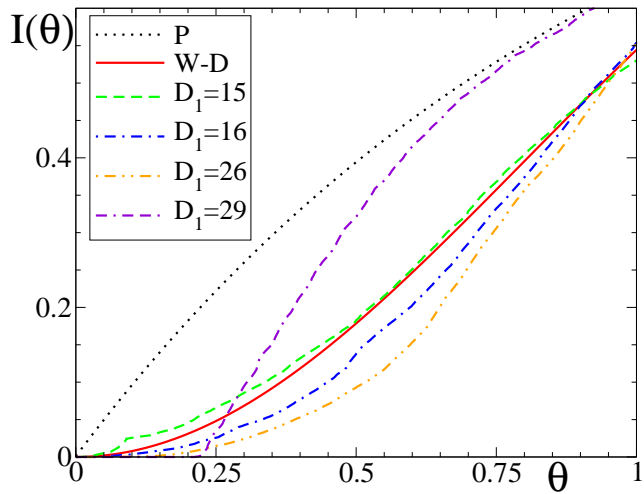}
\caption{(Color online) Cumulative level spacing $I(\theta)$ of the AQMBM averaged in $k$ 
($\theta$ is in units of the mean level spacing $2\pi/D$). This is shown for the Poisson 
and Wigner--Dyson distributions and for the AQMBMs displayed on Fig. \ref{fig:specD30dk} 
($D=30$; $D_1=15$, $D_1=16$, $D_1=26$ and $D_1=29$). See inset for references.
}
\label{fig:Chaos}
\end{center}
\end{figure}

We have analyzed the cumulative level spacing distribution of the AQMBM 
averaged in $k$, 
\begin{equation}
I(\theta)=\int dk/(2\pi)\int_0^\theta d\theta^\prime P(\theta^\prime),
\end{equation}
where $P(\theta)$ corresponds to the level spacing distribution. The results 
are shown in Fig. \ref{fig:Chaos}. We have taken the phase $\theta$ 
normalized by the mean level spacing $2\pi/D$. It becomes clear that the behavior 
of the case of the last panel in Fig. \ref{fig:specD30dk} ($D_1=29$) is completely 
different from the rest, confirming our previous conclusions. In fact, it is very 
close to the Poisson distribution, which corresponds to integrable or regular 
systems. Level repulsion is also evident 
since for small $\theta$ values, the curve corresponding to the AQMBM levels 
shows its main difference with respect to the Poisson one.
The other cases are very close to the Wigner--Dyson shape (CUE), that corresponds 
to the typical behavior of chaotic systems. It has to be underlined the very 
good agreement we have found for the symmetrical case. We can conclude that 
the quasi-regular behavior of the most asymmetric cases, i.e. 
the one we show for $D_1=29$ and the one for $D_1=28$ which is very similar to it, 
is highly anomalous. This is in close relation to the exceptional current values found for 
$D_1=D-1$ and $D_1=D-2$ in Fig. \ref{fig:c1}.

\section{Current generation}

In order to understand the origin of the directed current we have analyzed the 
classical and quantum phase space distributions for given initial conditions, 
as a function of time. We have studied them for short times and a 
Hilbert space dimension $D\leq80$, which is of the order of the Hilbert space dimensions 
of the cells we have used in obtaining the results of Section III. 
The choice of these evolution times and dimensions is suitable since it makes the phase 
space representations more clear and the departure of the quantum distributions 
from the classical behavior is already present. In fact, for our system this time 
can be extremely short, as we will see in the following. Then, although 
the asymptotic limit of the current $J_\infty$ is still far from being reached, 
the mechanisms that give rise to the current can be seen.

\begin{figure}[t!]
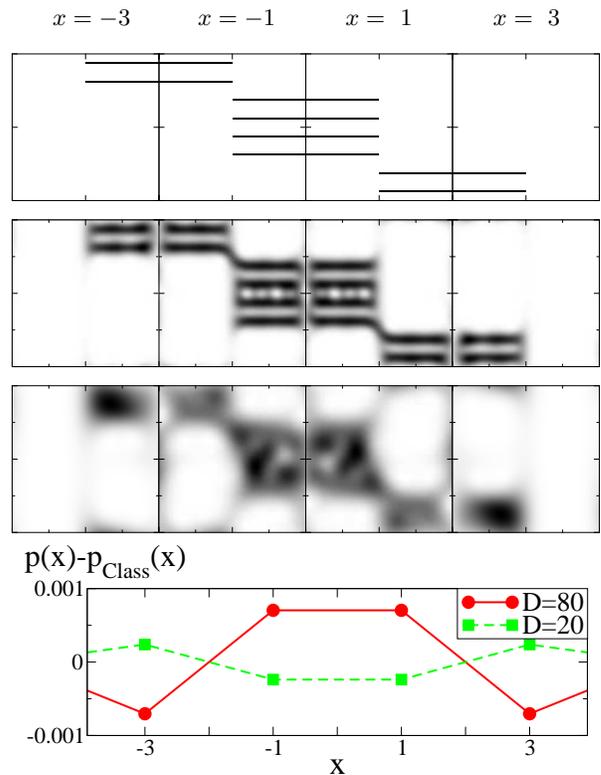

\begin{center}
$x=-3$\hspace{0.05\textwidth}$x=-1$\hspace{0.052\textwidth}$x=\ 1$\hspace{0.06\textwidth}$x=\ 3$
 \includegraphics[width=0.11\textwidth,angle=-90]{6a.ps}

\vspace{0.2cm}
 \includegraphics[width=0.11\textwidth,angle=-90]{6b.ps}

\vspace{0.2cm}
 \includegraphics[width=0.11\textwidth,angle=-90]{6c.ps}

\vspace{0.1cm}
 \includegraphics[width=0.42\textwidth,angle=0]{6d.eps}

\caption{(Color online) In the top panel we show the phase space of the classical Multibaker map with $s=0.5$ for a momentum centered 
strip of width $\delta p=0.1$ evolved 3 times. Only the sites with $x=-3,-1,1,3$ are shown. In the middle top and bottom 
panels the Husimi function is shown for the quantum version of the map 
($D=80$ and $D=20$, respectively). Finally, in the bottom panel we show the difference $p(x,t)-p_{class}(x,t)$ 
(see main text for details).}
\label{fig:Current1}
\end{center}
\end{figure}

\begin{figure}[t!]
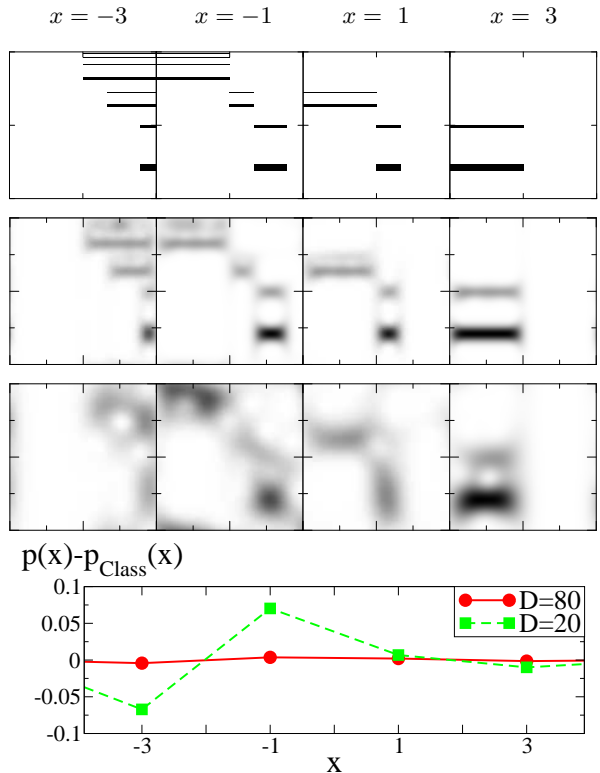

\begin{center}
$x=-3$\hspace{0.05\textwidth}$x=-1$\hspace{0.052\textwidth}$x=\ 1$\hspace{0.06\textwidth}$x=\ 3$
 \includegraphics[width=0.11\textwidth,angle=-90]{7a.ps}

\vspace{0.2cm}
 \includegraphics[width=0.11\textwidth,angle=-90]{7b.ps}

\vspace{0.2cm}
 \includegraphics[width=0.11\textwidth,angle=-90]{7c.ps}

\vspace{0.1cm}
 \includegraphics[width=0.42\textwidth,angle=0]{7d.eps}

\caption{(Color online) In the top panel we show the phase space of the classical Multibaker map with $s=0.75$ for a momentum centered 
strip of width $\delta p=0.1$ evolved 3 times. Only the sites with $x=-3,-1,1,3$ are shown. In the middle top and bottom 
panels the Husimi function is shown for the quantum version of the map 
($D=80$ and $D=20$, respectively). Finally, in the bottom panel we show the difference $p(x,t)-p_{class}(x,t)$ 
(see main text for details).}
\label{fig:Current3}
\end{center}
\end{figure}

An initial distribution corresponding to a momentum centered strip of width 
$\delta p=0.1$ and its quantum analogues have been evolved up to three time steps 
of the map. Results for $s=0.5$ are displayed in 
Fig. \ref{fig:Current1}, while the ones for $s=0.75$ are shown in Fig. \ref{fig:Current3}. 
In the top panels of both Figures we can see the classical distribution 
corresponding to the cells at lattice positions $x=-3,-1,1,3$, given that for $x=-2,0,2$ 
they are empty (this is a result of the translation operator and the initial 
conditions choice). In the middle ($D=80$) and bottom ($D=20$) panels we show the 
corresponding Husimi distributions, taking quantum initial conditions in the same 
way as in Section III. 
Finally, in the bottom panels we can find the probability distribution 
difference given by $p(x,t)-p_{class}(x,t)$.

\begin{figure}[htp]
\begin{center}
 \includegraphics[width=0.47\textwidth]{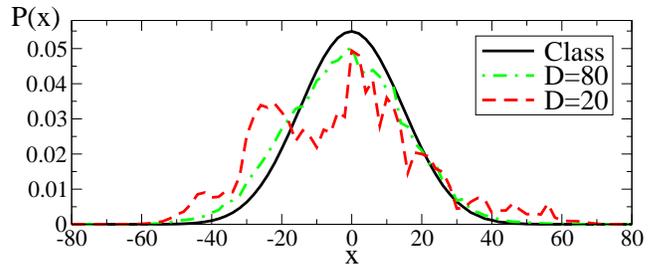}
\caption{(Color online) Classical $p_{class}(x,t)$ (solid black line) and quantum $p(x,t)$ (dot-dashed green line
 for $D=80$ and dashed red line for $D=20$) distributions, taking $s=0.75$ and $t=80$.}
\label{fig:pxt}
\end{center}
\end{figure}

By comparing both Figures we can immediately notice that the classical distribution 
$p_{class}(x,t)$ for $s=0.5$ keeps its initial symmetry. The quantum distributions in 
both cases considered 
also keep it. But for $s=0.75$ the situation changes. Now, the classical probability is not 
symmetrical but it is still balanced with respect to the origin (a given distribution is balanced 
if $<x>=0$). This asymmetry is also present in the quantum case, but the balance of the distribution 
is broken due to interference effects. 
In fact, if we look at the lower panel of Fig. \ref{fig:Current1} we can see that the quantum and 
classical distributions have almost equal weights in each cell (apart from quantum fluctuations).
But the lower panel of Fig. \ref{fig:Current3} clearly shows that for the $D=20$ case, the imbalance in 
the $p(x,t)$ distribution is already present. For $D=80$ we still have a close quantum-classical 
correspondence for this short evolution time. This fact underlines the fundamental role 
that quantum effects play in the net current appearance. It is clear that at times of the order 
of the Ehrenfest time ($t \sim \log_2{D}$) the imbalance starts to build up. This imbalance evolves 
in time shaping the $p(x,t)$ distribution. At the order of the Heisenberg 
time (which in this case corresponds to $t \sim D$) the asymptotic current is reached. We show the shape of 
$p_{class}(x,t)$ and $p(x,t)$ for the cases $D=20$ and $D=80$ in Fig. \ref{fig:pxt}, where we have 
taken $s=0.75$ and $t=80$. This illustrates how the probability distribution behaves at longer 
times.

\section{Conclusions}

In this work we have studied a recently introduced model for purely quantum directed 
transport, which shows a finite asymptotic current. We have analyzed the way in which 
the net transport appears by studying the classical and quantum phase space distributions 
for short times, showing the results for $t=3$. In the symmetric case $s=0.5$, the 
classical and quantum distributions retain the symmetry around $x=0$ and therefore 
both currents are forbidden. In the $s \neq 0$ case both distributions are asymmetric. 
The classical one is always balanced ($\langle x \rangle = 0$), while the quantum one 
develops imbalances leading to the appearance of a net current. This is clearly a 
purely quantum effect due to interferences. 

We have also studied several 
features of this phenomenon, in particular the dependence on the asymmetry parameter 
and the value of $h$. We could notice a marked dependence of the 
$J_{\infty}$ behavior on the values of $s$. In fact we observe a faster vanishing 
of the transport for $s<0.7$ both as $h \rightarrow 0$ and as the width of the initial conditions 
$\delta p \rightarrow 1$. We have found that for the higher values of $s$ the spectrum behavior 
approaches that of an integrable system (nevertheless with notable discrepancies, 
specially for small level spacings since no degeneracies are present).

We would like to mention that the mechanisms behind the current generation in our system 
are different from previously studied quantum ratchet accelerators \cite{BECratchets,coherentControl},
where there is a ballistic energy growth. Here, there is no need to control this 
effect since the asymptotic current is finite.

\begin{acknowledgments}
Partial support by ANPCyT and CONICET is
gratefully acknowledged.
\end{acknowledgments}

\end{document}